\documentclass[conference]{IEEEtran}
\IEEEoverridecommandlockouts
\usepackage{cite}
\usepackage{amsmath,amssymb,amsfonts}
\usepackage{algorithmic}
\usepackage{graphicx}
\usepackage{textcomp}
\usepackage{xcolor}
\usepackage{hyperref}
\usepackage[a4paper, total={184mm,239mm}]{geometry}
\def\BibTeX{{\rm B\kern-.05em{\sc i\kern-.025em b}\kern-.08em
    T\kern-.1667em\lower.7ex\hbox{E}\kern-.125emX}}
\begin{document}

\title{ChipLight: Cross-Layer Optimization of Chiplet Design with Optical Interconnects for LLM Training
}

\author{
\IEEEauthorblockN{Kangbo Bai}
\IEEEauthorblockA{\textit{School of Integrated Circuits} \\
\textit{Peking University}\\
Beijing, China \\
2301111810@stu.pku.edu.cn}
\and
\IEEEauthorblockN{Zhantong Zhu}
\IEEEauthorblockA{\textit{School of Integrated Circuits} \\
\textit{Peking University}\\
Beijing, China \\
2100012749@stu.pku.edu.cn}
\and
\IEEEauthorblockN{Yifan Ding}
\IEEEauthorblockA{\textit{School of Integrated Circuits} \\
\textit{Peking University}\\
Beijing, China \\
2401210237@stu.pku.edu.cn}
\and
\IEEEauthorblockN{Tianyu Jia}
\IEEEauthorblockA{\textit{School of Integrated Circuits} \\
\textit{Peking University}\\
Beijing, China \\
tianyuj@pku.edu.cn}
\vspace{-50pt}
}



\maketitle

\begin{abstract}
In large-scale distributed LLM training, communication between devices becomes the key performance bottleneck. Chiplet technology can integrate multiple dies into a package to scale-up node performance with higher bandwidth. Meanwhile, optical interconnect (OI) technology offers long-reach, high-bandwidth links, making it well suited for scale-out networks. 
The combination of these two technologies has the potential to overcome communication bottlenecks within and across packages. 
In this work, we present ChipLight, a cross-layer multi-objective design and optimization method for training clusters leveraging chiplet and OI. We first abstract an architecture model for such complex clusters, co-optimizing chiplet architecture, training parallel strategy, and OI network topology. Based on such models, we tailor the design space exploration flow by combining both black-box and white-box methodologies. Evaluated by our experimental results, ChipLight achieves significantly improved training efficiency and provides valuable design insights for the development of future training clusters.
\end{abstract}


\section{Introduction}

\begin{figure}[t]
  \centering
  \includegraphics[width=\linewidth]{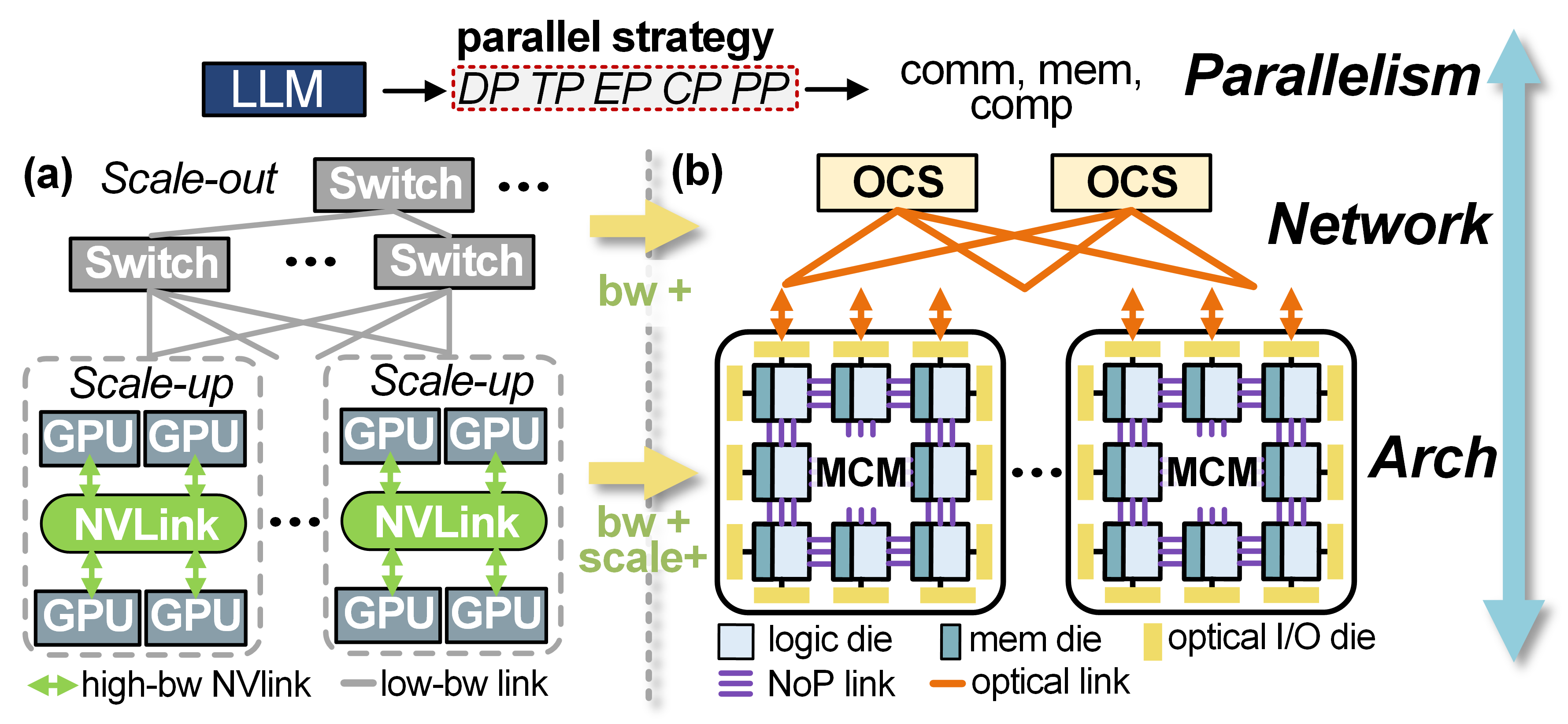}
  \vspace{-20 pt}
  \caption{LLM Training clusters and cross-layer optimization.}
  \label{fig:cross-layer}
  \vspace{-15 pt}
\end{figure}

Large language models (LLMs) have achieved widespread success across various applications. However, the growing model size and context length increase the cost and time of LLM training. 
As shown in Fig.~\ref{fig:cross-layer}(a), current LLM training typically relies on GPU clusters with scale-up nodes and scale-out networks. For example, GPU clusters can employ high-bandwidth NVLinks (900GB/s) for scale-up, and lower-bandwidth InfiniBand (IB, 60GB/s) links for scale-out \cite{naumov2020deep,deepseek}.
Prior research \cite{duan2024efficient,ub-mesh,mfabric} reveals that communication is the major performance bottleneck in large-scale LLM training. 
Emerging technologies hold promise for overcoming the “communication wall” for both scale-up and scale-out,  while also requiring novel architecture design and optimizations.

First, chiplet integration is a promising technology for node scale-up with higher bandwidth at larger scale. As shown in Fig.~\ref{fig:cross-layer}(b), chiplet is a technique using advanced packaging \cite{cowos,EMIB} to integrate multiple dies into a single package and form a multi-chiplet module (MCM). Chiplet not only reduces the single die area to improve manufacturing yield, but also offers much higher intra-package die-to-die bandwidths, e.g., up to 658 GB/s/mm \cite{chiplet_d2d} for high-bandwidth domains (HBD). Chiplet has already been adopted by industrial products, e.g., AMD MI300X \cite{MI300} integrates 8 chiplets. Attempts \cite{dojo,PD-aware} were also made to achieve larger integration scales.

Second, optical interconnect (OI) is a key technique for higher-bandwidth scale-out networks by replacing electrical interconnects with optical circuit switch (OCS) and links. 
As shown in Fig.~\ref{fig:cross-layer}(b), OI is critical, especially for MCM-based clusters, to avoid the benefits of on-package HBD being eliminated by low-bandwidth inter-MCM links. Co-packaged optics (CPO) can integrate optical modules within the package, achieving bandwidth density up to 128 GB/s/mm \cite{fathololoumi20244}. OI has already been adopted in Google TPUs \cite{tpu-v4} and will be employed in the upcoming Nvidia GPUs \cite{silicon_photonics}. 

We believe that the combination of chiplet and OI is a promising approach for next-generation LLM training clusters. However, there is a lack of research on the design, optimization, and modeling of such systems.
Prior studies have explored standalone MCM for inference tasks \cite{simba,cai2024gemini,sambanova,PD-aware,WSC_LLM} or solely on optical networks \cite{topoopt} for DNN tasks. Recently, RailX \cite{feng2025railx} explored a specific network design integrating MCMs and OI, but it lacks discussion on optimization of cluster architecture design and end-to-end evaluation of LLM training deployment. 


In this work, we propose ChipLight, a cross-layer design and optimization framework for LLM training clusters leveraging chiplet and OI techniques. 
Through cross-layer innovations, ChipLight can effectively exploit MCMs and OI for large-scale LLM training.
At the architecture level, the new design flexibility of MCMs is explored, including the optimization of single die area, integration scale, resource configuration of memory and OI ports. 
At the network level, the characteristics of training traffic is coordinated with the properties of MCM and OI for optimized scale-out networks. 
It is observed that the traffic distribution of LLM training exhibits both spatial and temporal traits, with the latter often neglected in prior optimizations \cite{topoopt, cosmic}.
A network model is developed to abstract the complex and unexplored topology design space, with the optimization flow decoupling the physical and logical topologies. 
At the parallelism level, parallel strategies are also adapted to the novel MCM and OI interconnects. 
After cross-layer optimization, clusters with MCM and OI showcase 19.58$\times$ throughput gains over conventional GPUs. Our optimization achieves a 41\% performance improvement over the latest network design \cite{feng2025railx} at a similar cost.


The contributions of this work are summarized as follows:

\begin{itemize}
\item We perform an in-depth analysis of LLM training traffic, leveraging its characteristics to design networks based on MCM and OI to enhance communication efficiency.

\item A design model of MCM- and OI-based network is developed to abstract the complex optimization space and to highlight key design parameters from MCM architecture, parallelism, and network topology.

\item A cross-layer multi-objective optimization method is proposed to perform effective explorations of the design for training clusters based on chiplet and OI. MCM architecture, network topology, and parallel strategy are co-optimized.  

\item 
After thorough optimization, we evaluate how chiplet and OI technologies enhance LLM training clusters, providing key insights for designing such systems.
\end{itemize}
\section{Background}
\subsection{Chiplet and Optical Interconnect}
\label{sec:network}
Chiplet has emerged as a crucial technology to scale-up performance and reduce costs. 
By leveraging advanced packaging technology, diverse dies are integrated on a substrate \cite{epyc} or interposer \cite{cowos} and connected by the network-on-package (NoP) to form an MCM. 
The NoP offers high bandwidth, low power, and low latency transfer compared to off-package interconnects. For example, novel die-to-die (D2D) interface achieves 658 GB/s/mm bandwidth at 0.29 pJ/b \cite{chiplet_d2d}. Chiplet also reduces costs by decreasing the single-die area and improving the manufacturing yield. 
The chiplet technology introduces greater flexibility in hardware design, and new design considerations have also been raised: For LLM training, what scale is required for a single chiplet and an MCM package? Should hardware configuration be adapted differently from GPUs? We aim to answer these questions.

To scale-out the cluster, optical interconnects are supplanting electrical links with higher bandwidth. With the recent CPO \cite{cpo1,cpo2,cpo3,cpo4} technology, optical I/O dies are integrated at the package edge for signal conversion between logic dies and external optical links. The optical links are connected with optical circuit switches (OCS). 
Our work focuses on micro-electrical mechanical switches (MEMS) OCS, which is currently the most promising pathway due to its high port count and technical maturity, and has been deployed in Google TPUs \cite{tpu-v4}. However, because of millisecond-scale switching latency, MEMS OCS needs one-time configuration of connections before training starts, or deterministic reconfiguration during non-communication phases. How can we build efficient OI networks, and what gains can they provide? We aim to answer these questions.

\subsection{Network Topology}

\begin{figure}[t]
  \centering
  \includegraphics[width=\linewidth]{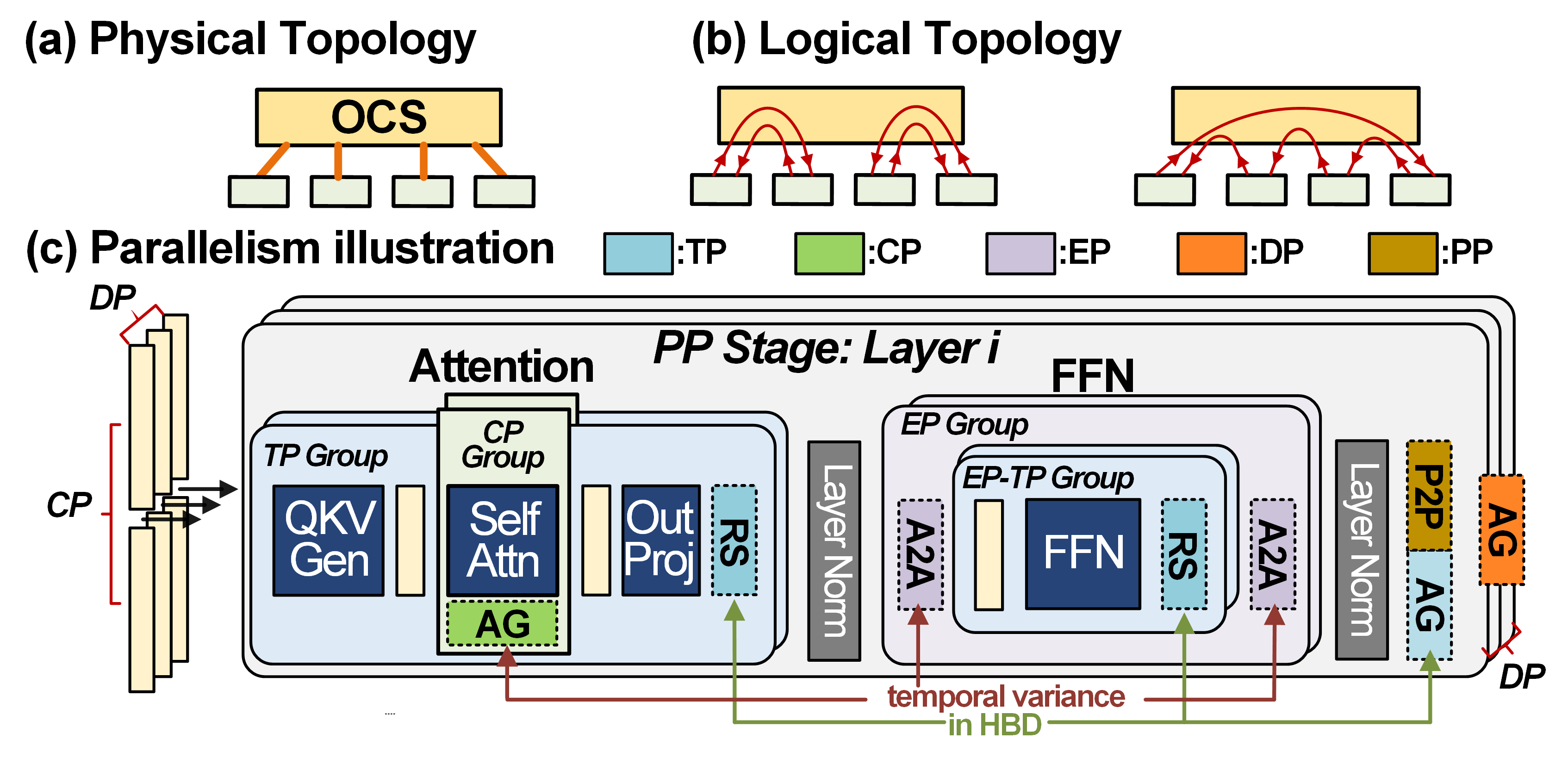}
  \vspace{-20 pt}
  \caption{Parallelism and corresponding communications in LLM training.}
  \label{fig:parallel}
  \vspace{-15 pt}
\end{figure}

As shown in Fig.~\ref{fig:parallel}(a), the network topology has two aspects: \textit{physical topology} and \textit{logical topology}. \textit{Physical topology} describes the actual wiring between devices and switches, and the \textit{logical topology} is the network structure upon which communication is implemented. Logical topology is derived from the physical topology by configuring switch connections and pruning idle links. Owing to the switch reconfiguration, a physical topology can support multiple logical topologies. As the example in Fig.~\ref{fig:parallel}(a), four devices are physically connected to an OCS, which can logically configure the devices as either two rings of diameter 2 or a single ring of diameter 4, as shown in Fig.~\ref{fig:parallel}(b). 
For LLM training, TPUv4 \cite{tpu-v4} employs a 3D-torus topology with OCS.
RailX \cite{feng2025railx} combines Chiplet and OI, which employs a HammingMesh-like physical topology and can be configured as a high-dimensional logical topology.  

\subsection{Parallelism in LLM Training}

Large-scale LLM training employs diverse parallelisms to distribute the workloads across numerous devices, e.g., GPUs or chiplets.  Fig.~\ref{fig:parallel}(b) illustrates an example of computation and communication for a Transformer layer under hybrid parallelisms, which is commonly supported by mainstream training frameworks \cite{megatron, deepspeed-TED}.

\textbf{Tensor parallelism} (TP) splits weight matrices and distributes shards across devices. Reduce-scatter (RS) and all-gather (AG) are required to aggregate the results.

\textbf{Data parallelism} (DP) splits data samples across devices, each of which keeps a model replica and processes data independently. 
All-gather (AG) communication is introduced during backward propagation to synchronize gradients.

\textbf{Pipeline parallelism} (PP) divides the model's layers into stages across devices in a pipelined manner, introducing peer-to-peer (P2P) send between adjacent stages.

\textbf{Context Parallelism} (CP) splits sequences across devices, completing attention computation through ring-attention \cite{ring_attention} to enable long context training, and all-gathering (AG) communication (green block in Fig.~\ref{fig:parallel}) is required in the attention.

\textbf{Expert Parallelism} (EP) is specialized for prevalent MoE LLMs \cite{deepseek,qwen3}, distributes FFN experts across devices. Before and after the FFN, all-to-all (A2A) communications are required to dispatch / recombine tokens to / from the devices hosting the corresponding experts.
\section{Profiling and Motivations}\label{sec:analysis}


\subsection{Profiling Analysis}\label{sec:volume}

\begin{figure}[t]
  \centering
  \includegraphics[width=\linewidth]{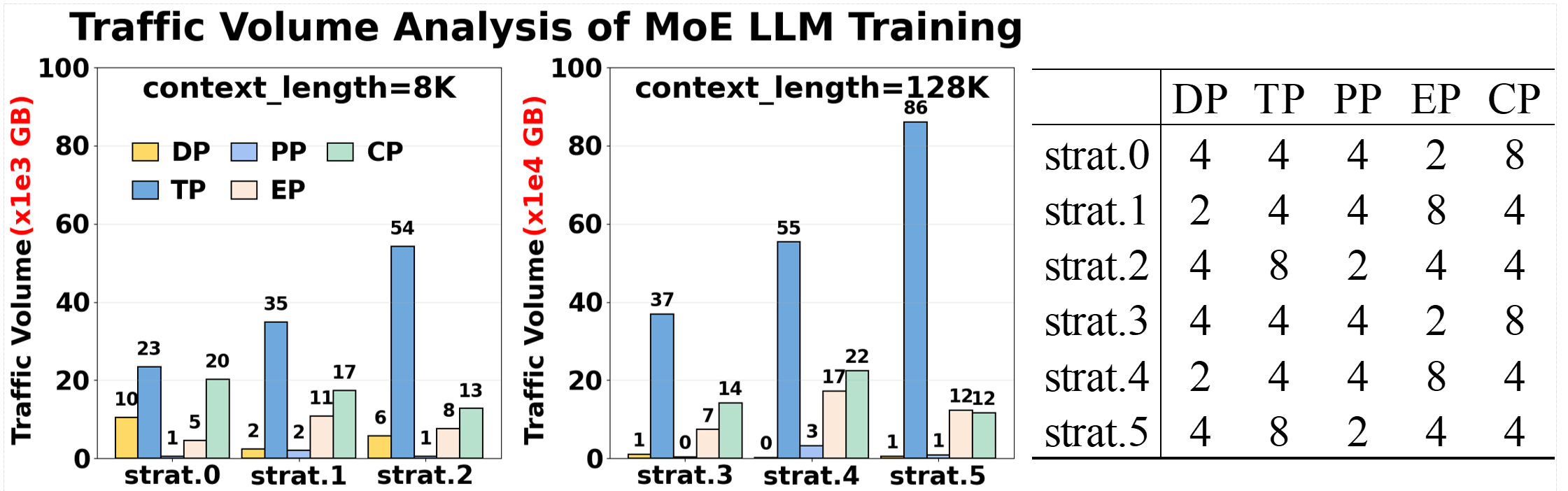}
  \vspace{-15 pt}
  \caption{Traffic volume of LLM training in different cases.}
  \label{fig:traffic_analysis}
  \vspace{-15 pt}
\end{figure}

 

We profile the training traffic volume for Qwen3-235B-A22B \cite{qwen3} on a total of 1024 devices (e.g., GPUs or chiplets), using ASTRA-sim \cite{astra-sim} simulator to derive generalizable observations for subsequent optimizations. The collective communication is based on the ring algorithm. 
Fig.~\ref{fig:traffic_analysis} shows the parallelism-wise traffic volumes under different parallel strategies (listed in the right table) and context length. 

Based on Fig.~\ref{fig:traffic_analysis}, we have concluded the \textbf{Observation 1: the parallel strategy significantly influences traffic volume, while it generally follows \textit{TP$>$(CP,EP)$>$(DP,PP)}}.
The relative order of the two parallelisms in parentheses also varies with the context length. Therefore, suitable parallel strategies are crucial, and networks with parallelism group mapping should be tailored to traffic volume.
Naturally, a TP group should be mapped within one GPU or MCM to confine the majority of traffic to on-package HBD. 
For chiplet node, depending on the MCM scale and parallel strategy, a parallelism other than TP may also be mapped within MCMs, while traffic of other parallelisms must traverse lower-bandwidth inter-MCM links. 
The current InfiniBand only offers 60GB/s bandwidth \cite{deepseek}, presenting a tens-of-times gap compared to TB/s-scale NoP bandwidth. However, as shown in Fig.~\ref{fig:traffic_analysis}, the CP and EP traffic is not much lower than TP. We have the \textbf{Observation 2: electronic links negate the benefits of on-package HBD due to traffic–bandwidth mismatch}.
In contrast, OI can increase inter-MCM bandwidth to 400GB/s or 128GB/s/mm \cite{mehta2024ai,fathololoumi20244}, significantly reducing the gap with NoP. 
\subsection{Traffic Distribution Analysis}\label{sec:traffic}
\begin{figure}[t]
  \centering
  \includegraphics[width=\linewidth]{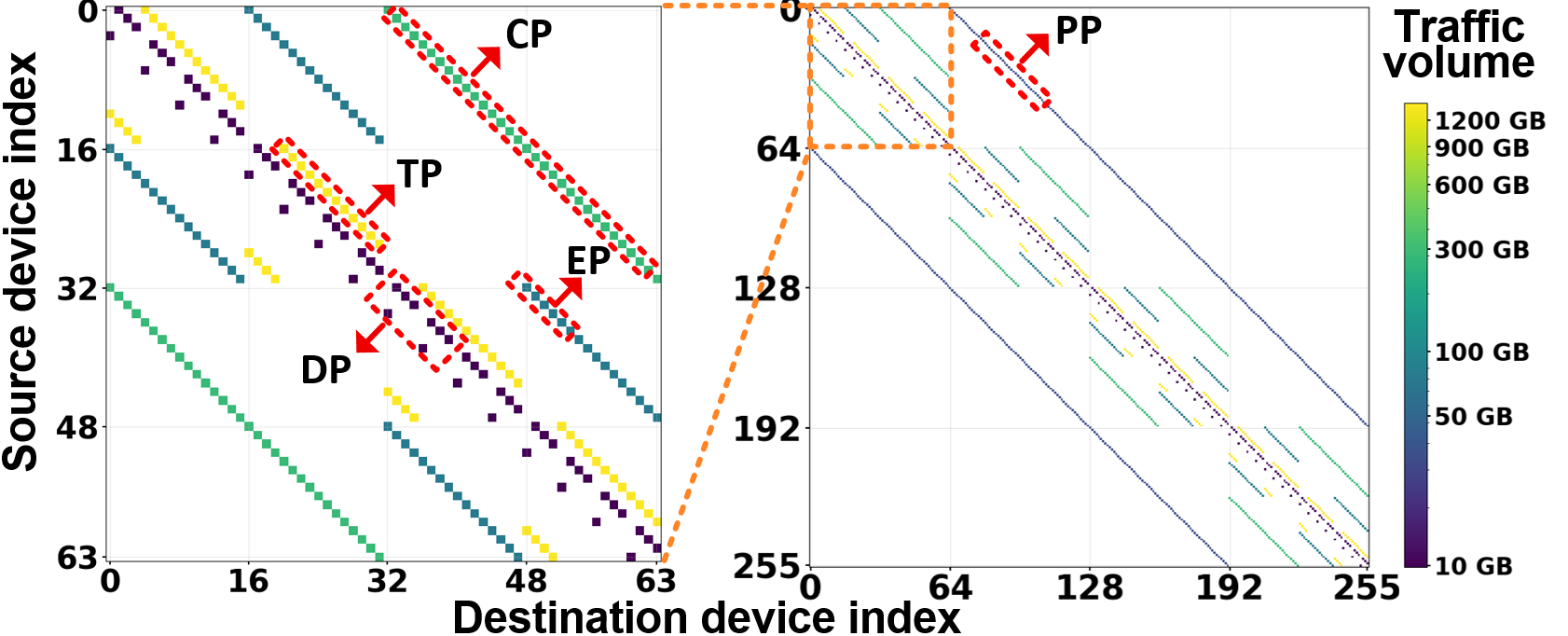}
  \vspace{-15 pt}
  \caption{Traffic distribution heatmap of LLM training.}
  \label{fig:traffic_heatmap}
  \vspace{-15 pt}
\end{figure}

In this section, we analyze the spatial distribution of training traffic for a ring topology, in which traffic is mainly concentrated on a small subset of links between neighbor devices. 
Fig.~\ref{fig:traffic_heatmap} shows a traffic heatmap with indices of source and destination devices on x- and y-axes, and each grid’s color indicates the corresponding training traffic volume. It is evident that the training traffic is highly sparse and regular. 
Traffic is confined within parallel groups, and due to the ring communication algorithm, each device typically communicates only with a few devices that are adjacent in the ring.
This sparsity leaves many links idle in conventional all-to-all networks in data centers. 
Additionally, the traffic volumes vary greatly, reflecting uneven distribution. Based on the above characteristics, we have the \textbf{Observation 3: training traffic is spatially sparse and link resource allocation should be guided by traffic distribution.}

We also analyze the temporal distribution of training traffic, in which communication of various parallelisms does not occur continuously during the training. As shown in Fig.~\ref{fig:parallel}, CP and EP traffic occur in the attention and FFN operators respectively, exhibiting temporal variance due to intervening output\_proj and layernorm computations. We have the \textbf{observation 4: the temporal traffic variance allows dynamic reuse of bandwidth resources}, which has been neglected in prior optimizations \cite{topoopt,cosmic}. If two independent parts of links are allocated to the CP and EP traffic, one part will remain idle while the other is active. Conversely, reconfiguring the network via OCS between the interval of CP and EP traffic allows links to be reused for both traffic, improving network efficiency. Opportunities for link reuse also exist among CP, DP, and PP. But as mentioned in Sec.~\ref{sec:volume}, CP and EP traffic typically dominate the inter-MCM network. Therefore, reuse between these two parallelisms is most beneficial. To support dynamic link reuse, the OCS switching latency should be smaller than the traffic interval, which is satisfied in practice. 

\section{Cross-Layer Optimization for Training Cluster with Chiplet and OI}
\subsection{Model of Chiplet and OI Network}

Characterizing the design points and defining the optimization space of MCM- and OI-based training clusters is non-trivial due to their complex structure. To explore a wide range of promising and crucial design possibilities under controllable optimization complexity, we abstract a cluster model for subsequent optimizations.

\begin{figure}[t]
  \centering
  \includegraphics[width=\linewidth]{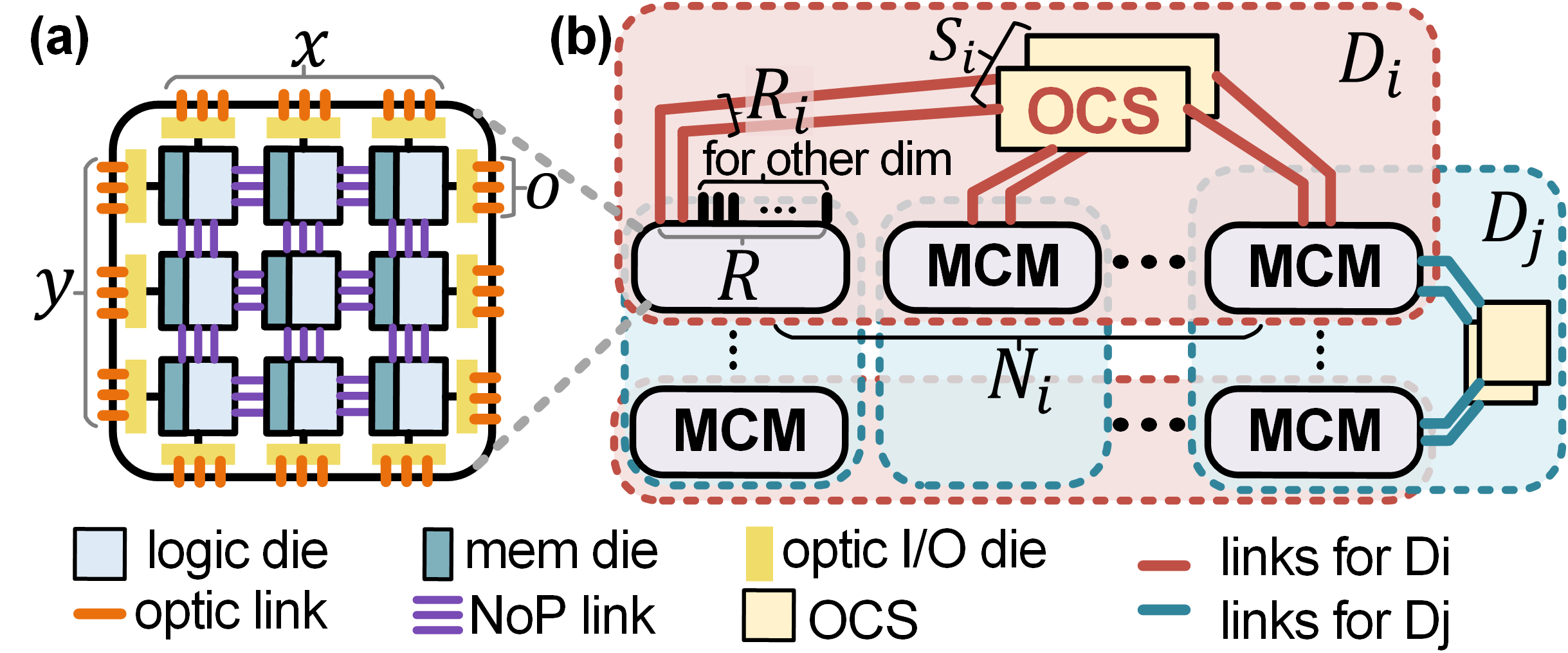}
  \vspace{-20 pt}
  \caption{Our cluster model for training cluster with MCM and OI.}
  \label{fig:template}
  \vspace{-15 pt}
\end{figure}

Fig.~\ref{fig:template}(a) shows the model of an MCM. 
To explore the design flexibility of MCM, we set the total compute performance $C$ of the cluster as the input constant rather than the chiplet count. $C$ is split into $N$ MCMs. An MCM is further split into $x\times y$ logic dies, each coupled with $m$ memory dies. 
Optical I/O dies are integrated at the edge of the MCM package, each providing $o$ optical links. An MCM has a total of $L=2\times (x+y) \times o$ links for external connectivity. 
The maximum of $o$ is proportional to the edge length of the logic die. 
A larger $o$ requires more OCS, significantly increasing cost. 
The design variables $N$, $x$, $y$, $m$, $o$ define the MCM architecture. We assume that NoP employs mesh topology with bandwidth $B_{p}$, which scales linearly with the length of D2D. The edges of logic die are shared by the D2D link interfaces and memory dies, leading to a trade-off between $m$ and $B_{p}$.

Fig.~\ref{fig:template}(b) shows the OI network model. It is challenging to describe complex and diverse physical topologies using quantitative variables. Inspired by previous research \cite{tpu-v4,feng2025railx}, we abstract diverse physical topologies into a rail-based model, which consists of multiple \textit{rail dimensions}. Fig.~\ref{fig:template}(b) illustrates an example with two rail dimensions, i.e., $D_{i}$ and $D_{j}$. A rail is a connection comprising multiple MCMs, where each MCM connects to an OCS via a single link. A rail dimension $D_{i}$ represents a set of connections, where $S_{i}$ OCSs are used to connect $N_{i}$ MCMs, occupying $R_{i}$ links on each MCM, and is denoted as $D_{i}=(N_{i},R_{i},S_{i})$.  
Each MCM connects $k_{i}$ links to the same OCS. So $S_{i}=\left \lfloor R_{i}/k_{i} \right \rfloor$, Fig.~\ref{fig:template}(b) shows the $k_{i}=1$ case. 
Each OCS has $P$ ports, and $k_{i} \times N_{i}\le P$ needs to be satisfied. Each MCM is involved in multiple rail dimensions, and their interweaving constitutes the entire network.
The MCM count satisfies $\prod N_{i} =N$, and OCS count  $S=\sum_{i}(\prod_{j\ne i}N_{j} \times S_{i})$. Due to symmetry, link allocation across $D_{i}$ is uniform for all MCMs and satisfies $\sum R_{i}=L$.  

Besides OI network, the parallelism values are also incorporated as variables in the model. Parallelism groups are mapped to intra-MCM HBD or inter-MCM rail dimensions. There is no one-to-one correspondence between rail dimension and parallelism, as multiple parallelisms can map to a single $D_{i}$. Since up to four inter-package parallelisms in LLM, $i\le 4$. 

The advantage of our model lies in comprehensive coverage of optimization space, allowing cross-stack exploration of design possibilities that have been ignored before. For example, RailX \cite{feng2025railx} and TPUv4 \cite{tpu-v4} can be viewed as specific cases that have two or three rail dimensions with uniformly distributed links. Prior optimization considered a single-dimensional network \cite{topoopt}, or simplified practical physical topology \cite{cosmic}. Multiple key design aspects of training clusters have been considered, including the scale of MCM and chiplet, and the allocation of compute, memory, and NoP.

\subsection{ChipLight Optimization Methodology}

\subsubsection{Optimization Workflow}
\begin{figure}[t]
  \centering
  \includegraphics[width=\linewidth]{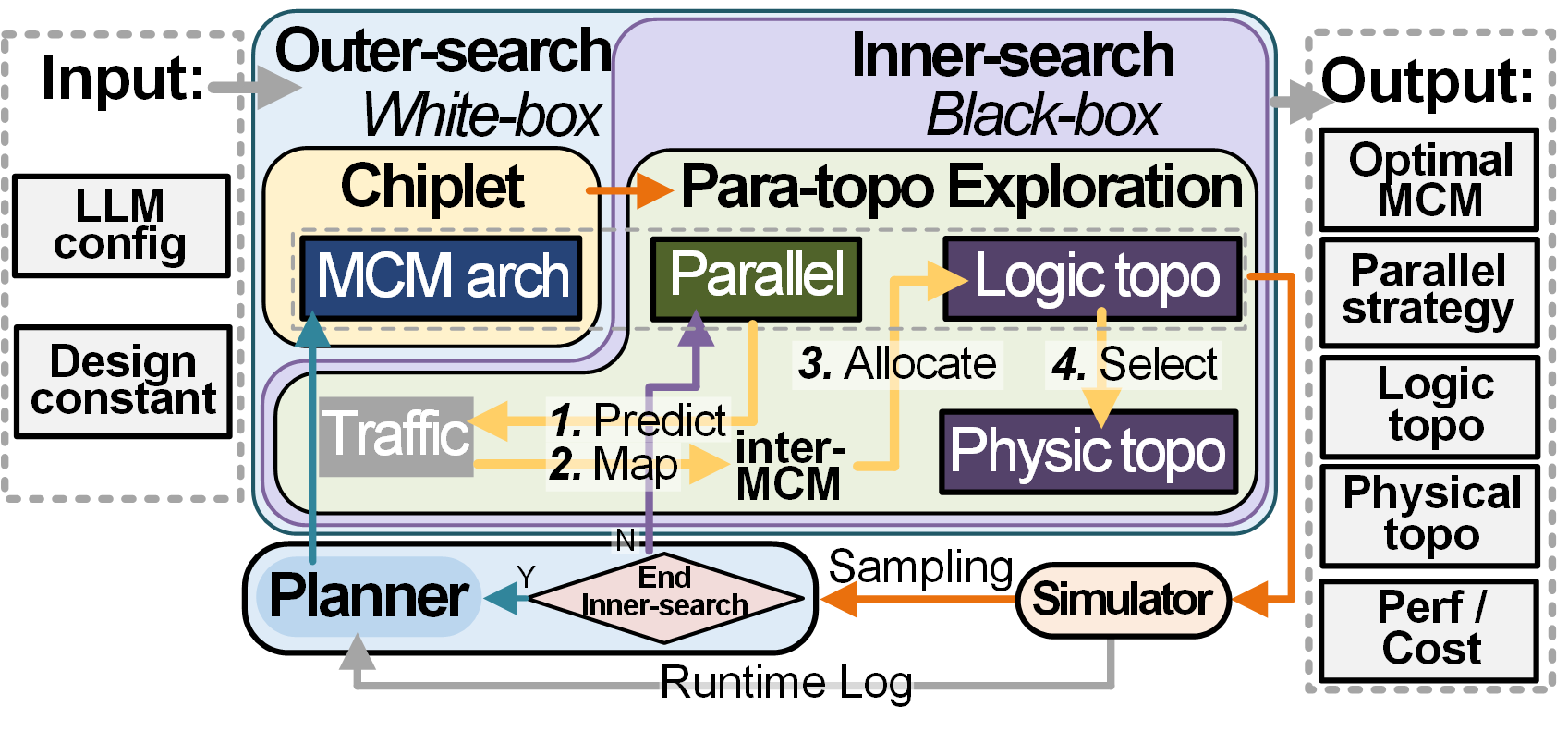}
  \vspace{-20 pt}
  \caption{Workflow of ChipLight cross-layer optimizations.}
  \label{fig:method}
  \vspace{-10 pt}
\end{figure}

ChipLight employs a cross-layer multi-objective method as explained in Fig.~\ref{fig:method}. The LLM configuration and other design constants are set as input. Since the MCM architecture defines the design space of parallel strategy and OI networks, 
we adopt a nested optimization flow: The outer-search explores the MCM architecture, while for each MCM sample, the inner-search optimizes the parallel strategy and OI network topology, i.e. the para-topo explorations. 
Each sampled design point is evaluated by simulator. 
For the next iteration, the planner decides whether to continue the inner search or switch to the outer-search. In the latter case, the new MCM sample is also determined. Through iterations, a performance-cost Pareto frontier is generated, representing the optimized MCM architecture, topology, and parallelization strategy. 
We combine black-box and white-box optimization methods for above flow, detailed as follows.

\subsubsection{Optimization for Para-Topo}\label{sec:para-topo}
\begin{figure}[t]
  \centering
  \includegraphics[width=\linewidth]{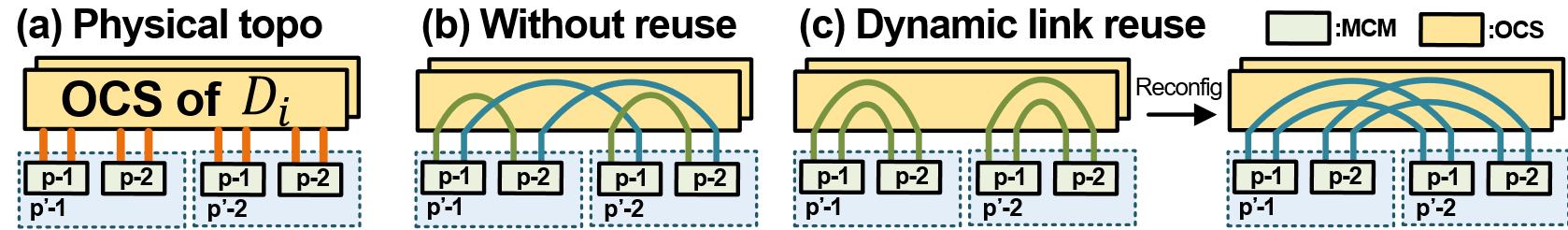}
  \vspace{-15 pt}
  \caption{Illustration of the dynamic link reuse.}
  \label{fig:dynamic}
  \vspace{-15 pt}
\end{figure}
In inner-search for parallelism and topology co-explorations, a straightforward flow first set the OI physical topology, then tailors the parallel strategy and logical topology accordingly.
However, such methods may converge to a local optimum due to the pre-fixed physical topology and have low search efficiency. In contrast, ChipLight follows a \textbf{parallel-centric} search flow.

As shown in Fig.~\ref{fig:method}, the parallelism values are set as optimization variables in the inner-search and sampled by a black-box algorithm, e.g., PRF \cite{prf}. First, the traffic volume is projected based on the determinacy of training traffic and is independent of the underlying network, similar as the results in Fig.~\ref{fig:traffic_heatmap}. 
Second, parallelism groups are mapped to intra-MCM or inter-MCM. The TP group and possibly one other group are mapped to intra-MCM, while the traffic of other parallelisms traverses the OI network.
Third, links of an MCM are allocated to the traffic of inter-MCM mapped parallelism. For parallelism $p$ with traffic volume $v_{p}$, each MCM assigns $l_{p}$ links to parallelism $p$ based on its traffic proportion, i.e., $l_{p}=\left \lfloor L*v_{i}/(\sum v(i))\right \rfloor$. An ideal logical topology can be generated in which all MCMs involved in an inter-MCM parallelism group form a ring. The ring size equals the value of this parallelism, and bandwidth scales with $l_{p}$. For EP, fully-connected (FC) topology can also be considered for A2A communications. The overall logical topology is a high-dimensional composition of local ring/FC of each parallelism, with all links being utilized and aligned to the spatial-temporal distribution of traffic. 
Finally, the physical topology is derived from the ideal logical topology. For each parallelism $p$, the ring of $p$ must be mapped to a rail dimension $D_{i}$, with $l_{p}=D_{i}$. If two parallelisms are mapped to the same $D_{i}$, the product of their values equals $N_{i}$. 
By enumerating all parallelism-rail mappings, feasible physical topologies can be identified, and the one with the fewest OCSs is selected. 

Dynamic link reuse can be supported by the above flow with minor modifications. If dynamic link reuse is applied between $p$ and $p'$, they must be mapped to the same $D_{i}$. In other words, the MCMs involved in $p$ and $p'$ must physically be connected to the OCS of $D_{i}$ as shown in Fig.~\ref{fig:dynamic}(a). Fig.~\ref{fig:dynamic}(b) illustrates the case without link reuse, where each MCM uses a green link for $p$ traffic and a blue link for $p'$ traffic. Fig.~\ref{fig:dynamic}(c) shows the link reuse implementation, where each MCM employs two links for $p$; and through OCS reconfiguration, the two links are fully used for $p'$. Therefore, in link allocation, $p$ and $p'$ share the links with count:
\begin{equation}
\label{cp-ep}
l_{reuse}=\left \lfloor \frac{L*max(v_{p},v_{p'})}{\sum v_{others}+max(v_{p},v_{p'})}  \right \rfloor 
\end{equation}
During the physical topology selection, we impose the constraint that $p$ and $p'$ are mapped to the same rail dimension. 




\subsubsection{Optimization for Chiplets}
By leveraging the simulation runtime logs, we conducted an efficient outer-search for chiplet architecture. 
Outer-search sampling is costly since each sample introduces a full inner-search. Current optimizations \cite{cai2024gemini} often neglect the development of sampling strategies in outer-search. However, for highly flexible MCM architectures, exploring the vast search space with limited outer-sampling iterations necessitates efficient sampling methods.
Conventional black-box algorithms perform poorly with limited samples. 

To improve the search efficiency, additional information is required beyond the standalone final performance and cost metrics.
Simulation logs contain abundant information that reveals hardware bottlenecks and utilization. Resampling of MCM architecture occurs after an inner-search, when parallelism and network are near-optimal, and the bottleneck lies in the MCM architecture.
We implement a heuristic planner to identify the bottleneck by collecting key metrics in logs, such as compute resource utilization, memory consumption, and communication bottlenecks, and modify the MCM architecture to break bottlenecks or reduce redundancy. 
\section{Evaluations and Insights}\label{sec:evaluation}

\subsection{Evaluation Setup}
In our experiment, the design parameters of logic dies are based on H100 GPU \cite{h100}, while the memory die employs HBM3 \cite{hbm3}. The parameters for the chiplet are derived from \cite{chiplet_d2d,WSC_LLM}, while the CPO implementation is based on \cite{fathololoumi20244, mehta2024ai}. Target model is Qwen3-235B-A22B \cite{qwen3} with context length of 10k. We modified the open-source ASTRA-sim \cite{astra-sim} to serve as our evaluation simulator. The cost estimation is based on  \cite{chiplet_actuary,feng2025railx}.

\subsection{Breaking Communication Bound by Chiplet and OI}

\begin{figure}[t]
  \centering
  \includegraphics[width=\linewidth]{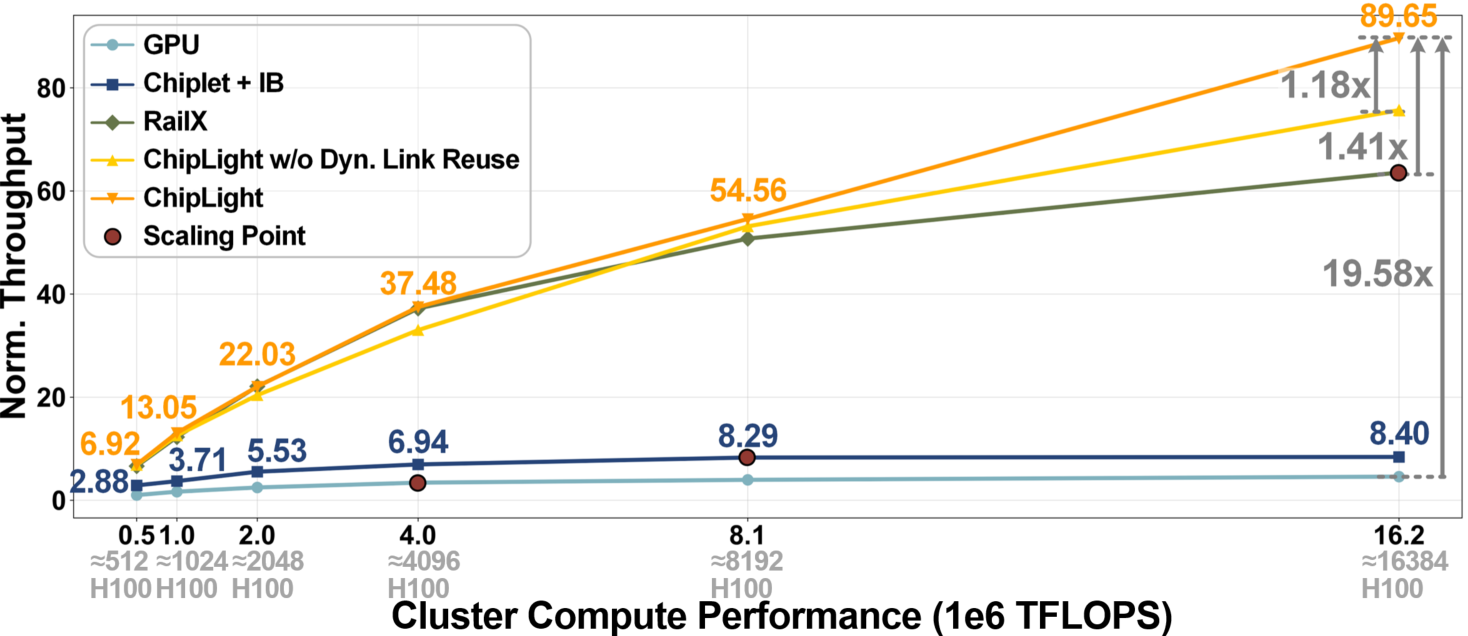}
  \vspace{-15 pt}
  \caption{Training throughput scaling across different clusters.}
  \label{fig:trend}
  \vspace{-15 pt}
\end{figure}

We first evaluate the throughput trends of the training cluster under different hardware choices.
Fig.~\ref{fig:trend} shows the end-to-end training throughput (y-axis) changes as the total compute performance $C$ (x-axis) increases. We apply our optimization to each experimental group to the extent permitted, exploiting their potential.
H100 \cite{h100} GPU clusters with clos network formed by NVLink and IB are used as the comparison baseline. Throughput of such clusters increases proportionally with $C$, or GPU counts, at small scales. 
However, the growth rate significantly slows down when $C$ exceeds 4$\times 10^{6}$TOPS, which we define as the scaling point. While the per-device computation load decreases proportionally with the number of devices, the communication traffic volume and allocated bandwidth resources remain relatively constant. Therefore, the training throughput encounters a communication wall at large scale.

Next, we evaluate the gains of chiplet and OI technology. To ensure a fair comparison, each logic die and its coupled HBM are consistent with the H100. Chiplet+IB employs NoP instead of NVLink for scale-up, while retaining electrical IB in scale-out network. As shown in Fig.~\ref{fig:trend}, Chiplet+IB demonstrates modest gains over GPU baseline and pushes the scaling point to about 8$\times 10^{6}$TOPS. However, it still encounters a clear boundary caused by IB. 
Both RailX and ChipLight combine chiplets with OI, the former builds on existing design \cite{feng2025railx}, and the latter is refined through our network optimization. Both show a clear throughput boost over GPU clusters, especially at large scales. The scaling point is pushed far out, indicating enhanced system scalability. Therefore, we obtain \textbf{insight 1: chiplet technology alone cannot fully overcome the communication bound, while its combination with OI greatly alleviates the communication bottleneck.}

In Fig.~\ref{fig:trend}, RailX and ChipLight are configured with the same per-edge OI ports $o$, resulting in similar total cost. We also apply dynamic link reuse in RailX. At small scales, throughput of RailX and ChipLight is comparable, as the OI network is not the bottleneck, reducing the necessity of optimization. However, at $C = 16\times 10^{6}$TFLOPS, ChipLight achieves 41\% improvement over RailX due to optimized network topology. We also evaluate the case without dynamic link reuse, which has a throughput drop of ~30\%. This leads to \textbf{insight 2: through optimizations on network topology and dynamic link reuse, ChipLight achieves better network efficiency with similar cost}.

\subsection{Explorations on Optimal Packaging and Single Die Scale}

With our ChipLight method, we can explore the optimal chiplet design scale for training clusters. 
In our experiment, the total compute performance of the cluster is fixed at $C = 8\times 10^{6}$TFLOPS, and the single logic die has the same specifications as H100 GPU. 
We swept the number of logic dies integrated on MCM (i.e., MCM\_scale) to explore the impact of package scale on cluster throughput and cost. 

\begin{figure}[t]
  \centering
  \includegraphics[width=\linewidth]{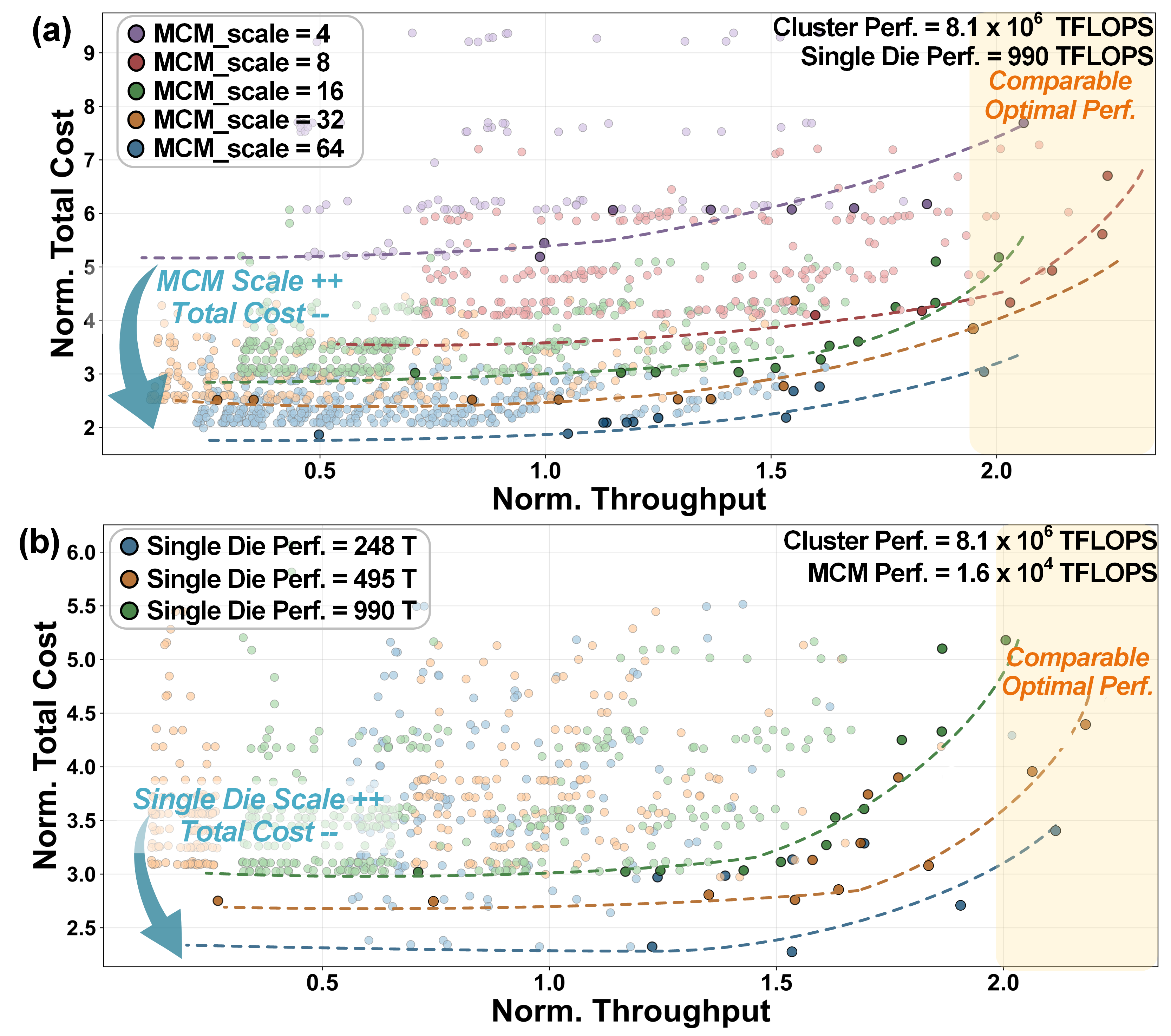}
  \vspace{-20 pt}
  \caption{Cost-throughput landscape for different MCM and single die scale.}
  \label{fig:pareto}
  \vspace{-15 pt}
\end{figure}

Fig.~\ref{fig:pareto}(a) presents the throughput-cost distribution of the cluster as the MCM\_scale ranges from 4 to 64, plotted in different colors. The Pareto fronts are marked with dashed lines. 
After our thorough optimization, clusters based on small-scale MCMs achieve comparable optimal throughput to those using large-scale MCMs, and even marginally surpass the latter in some cases. This is because OI narrows the bandwidth gap between HBD and the scale-out network, reducing the benefit of enlarging HBD. 
And in larger MCMs, there are more inner logic dies, reducing the per-die OI bandwidth. 
Furthermore, the mesh topology becomes less efficient in large-scale NoP. 
Under fixed total compute performance, it is observed that larger-scale chiplet integration shifts more interconnects from OI to NoP, requiring fewer OCS and consequently greatly lowering cost. 
Hence, we have \textbf{insight 3: small-scale MCMs with OI can match the performance of large-scale MCMs, while large-scale integration greatly reduces cluster cost.}

Chiplet technology has better manufacturing yield and lower cost.
However, smaller and more logic dies increase traffic volume, potentially leading to performance degradation. Fig.~\ref{fig:pareto}(b) shows the cost-throughput distribution of clusters with varying logic die scale. We maintain a constant compute performance of the entire cluster and an MCM for comparison. Fig.~\ref{fig:pareto}(b) indicates that reducing the performance of a logic die by half or even a quarter, i.e., cutting its area, does not lead to significant performance degradation. It is manifested as a downward shift of the Pareto frontier. The reason is that the high-bandwidth NoP can accommodate the traffic load growth caused by the increased count of logic dies. The total cluster cost decreases by approximately 23\% when the single die scale is reduced to one quarter. We have \textbf{insight 4: by co-optimization such as ChipLight, it is feasible to reduce the die area and cost while maintaining the cluster performance.}

\subsection{Explorations on Memory Resource and OI Links}

Beyond the logic die, memory and optical I/O dies also influence the cost and performance of training clusters. For such novel clusters with MCM and OI, the configuration of these resources also needs to be re-evaluated. 

\begin{figure}[t]
  \centering
  \includegraphics[width=\linewidth]{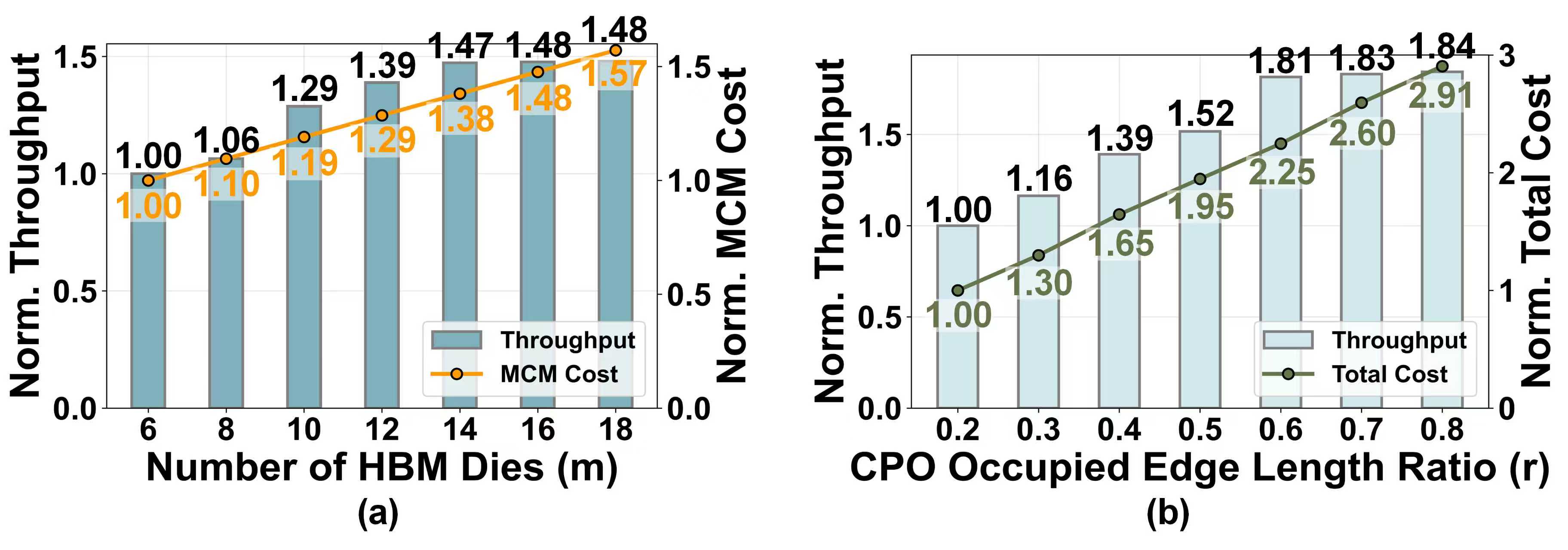}
  \vspace{-20 pt}
  \caption{Trade-off for memory and OI resource.}
  \label{fig:memOI}
  \vspace{-15 pt}
\end{figure}

Fig.~\ref{fig:memOI}(a) plots the trend of cluster cost (in line) and throughput (in bar) versus the number of memory dies ($m$) that are coupled with a logic die, while other hardware configurations are fixed. 
In practice, H100 is equipped with six HBM dies, providing about 3.3 TB/s of memory bandwidth.
For a logic die with compute performance equivalent to H100 GPU, increasing the number of attached HBM improves cluster throughput effectively, until $m$ is near 14, which indicates
more HBMs should be placed with GPUs.
The reason is straightforward. Collective communications, such as reduce-scatter, occur along with aggregation computation. 
After data is received, memory dies must perform at least one read and one write. 
Since the GPU memory bandwidth is multiple times that of NVLink, the latency of memory access can be overlapped in data transmission. 
For MCMs, the bandwidth of the NoP and the memory are comparable, the latter can become the bottleneck.
However, a large $m$ also increases the cost, as shown in Fig.~\ref{fig:memOI}(a). In our case, 14 HBMs is the optimal configuration for abundant bandwidth, but it is difficult for practical physical implementations.
So we have \textbf{insight 5: compared with GPUs, logic dies in MCMs require more HBMs and call for future higher bandwidth memory solutions}.

Fig.~\ref{fig:memOI}(b) shows the impact of cluster throughput and cost as more OI ports are integrated on a logic die. Current CPO achieves a bandwidth density of 128 GB/s/mm \cite{mehta2024ai}. However, during explorations, we observe that it is not optimal for logic die to dedicate the entire edge for CPO to maximize optical ports. We use $r$, the ratio of edge length that is occupied for CPO as the x-axis in Fig.~\ref{fig:memOI}(b), with the corresponding cluster cost and optimal throughput as the y-axis. Increasing $r$ indicates more optical links are utilized to boost bandwidth, and more OCS are required to support the connections. It is observed that the high cost of OCSs imposes a significant increase on the total cluster cost. As shown in the Fig.~\ref{fig:memOI}(b), the throughput growth becomes disproportionate to the additional cost when $r$ is beyond 0.6. This indicates that a balanced cost-performance configuration only utilizes over half of the maximum bandwidth of CPOs. Hence, we have \textbf{insight 6: the current bandwidth density of CPO is sufficient to meet interconnect demands, while the costly OCS limits the available OI bandwidth in practice}.


\section{Conclusion}

In this paper, we explore the design space of training clusters with chiplet technology and optical interconnect. We present ChipLight, a cross-layer optimization method with cluster models for MCM and OI. 
Compared to conventional GPU clusters, the optimized clusters with MCM and OI achieve a 19.58$\times$ improvement in training throughput under our explorations. 
Under the same technology and comparable cost, ChipLight achieves a 41\% throughput improvement compared to prior network design \cite{feng2025railx}.
Valuable design insights for cluster architectures have also been provided.

\section*{Acknowledgment}
This work was supported in part by NSFC No. 92464202.

\bibliographystyle{ieeetr}
\bibliography{refs}
\end{document}